\documentclass[prb,a4paper,twocolumn,showpacs,floatfix]{revtex4-1}

    \usepackage[utf8]{inputenc}
    \usepackage[T1]{fontenc}

    \usepackage{amssymb}
    \usepackage{amsmath}
    \usepackage{amsbsy}
    \usepackage{bm}
    \usepackage{graphicx}
 \usepackage{xcolor}

\begin{document}
\title{Spin waves scattering on a Bloch point}
\author{R. G. El\'ias$^1\email{gabriel.elias@usach.cl}$, V. L. Carvalho-Santos$^{1,2}$, A. S. N\'u\~nez$^3$, A. D. Verga$^4$}%

\affiliation{(1) Departamento de F\'isica, Universidad de Santiago de Chile, Avda. Ecuador 3493, Santiago, Chile.}
\affiliation{(2) Instituto Federal de Educa\c c\~ao, Ci\^encia e Tecnologia
Baiano - Campus
Senhor do Bonfim, \\Km 04 Estrada da Igara, 48970-000 Senhor do
Bonfim, Bahia, Brazil.}
\affiliation{(3) Departamento de F\'isica, Facultad de Ciencias F\'isicas y 
Matem\'aticas, Universidad de Chile, Casilla 487-3, Santiago, Chile}
\affiliation{(4) Universit\'e d'Aix-Marseille, IM2NP-CNRS, Campus St. J\'er\^ ome, Case 142, 13397 Marseille, France}

\date{\today}

\begin{abstract}
We show that, after a transformation, the dynamics of linear perturbations (spin waves) around a singular Bloch point soliton is formally equivalent to a quantum system of an electron in a magnetic monopole field. The analytical solution to this problem is known and allows us to find the spectrum and the scattering of a wave in a Bloch point field.
\end{abstract}


\maketitle

\section{Introduction}
Bloch points (BPs) are topological solitons found in three-dimensional magnets.  They have been observed or inferred in different contexts, such as in the transition region between Bloch lines embedded in Bloch walls\cite{Hubert-1998vn} and in numerical simulations of quasi-two-dimensional systems during the process of reversal of vortex cores.\cite{PhysRevB.67.094410, Hertel-2004ys, Hertel-2007jy} More recently, BPs have been identified as required sources and sinks for the unwinding of a skyrmion lattice.\cite{Milde31052013} Remarkably, there is also a recent experimental work in which a static BP was observed in cylindrical magnetic nanowires.\cite{PhysRevB.89.180405} The defining property of BPs is that in a closed surface around its center the direction of the magnetization field covers the whole solid angle an integer number of times. When the norm of the magnetization field is preserved this property turns into a topological protection (in which case the BP center is a singular point where ferromagnetic order is destroyed). Actually, BPs can be seen as the equivalent of two-dimensional Belavin-Polyakov solitons \cite{Belavin-1975xw} (skyrmions) if we fold the physical plane over the surface of a sphere by means of the stereographic projection. In this sense, the simplest BPs are solitons with unitary topological charge and it is for this reason that they are implicated in topological transitions where topological charge always changes by steps of $\pm1$. Even in strictly two-dimensional system we can observe the appearance of a BP-like configuration given by the superposition of two magnetic vortices (a vortex and an anti-vortex) with the same topological charge (Pontryagin invariant) but opposite vorticity. \cite{PhysRevB.58.8464,PhysRevB.65.134434,PhysRevB.89.134405} The control and manipulation of topological solitons (principally vortices and skyrmions) by means of electrical currents in the hope to find new alternatives for the information storage has relaunched in recent years the investigation on BPs. Another potential utilization of BPs is in the field of magnonics that pretends to manipulate magnetic solitons by means of the spin waves generated in the material,\cite{Han-2009lo,PhysRevLett.107.177207} preventing in this way the Joule effect produced by currents. In any case, the knowledge of spin waves behavior is of paramount importance to understand and to control BPs dynamics. 

Considering these facts it is worth to know the dynamical and stability properties of BPs in a ferromagnetic materials. For this purpose, in this paper we study the spin  waves (SWs) around a singular BP described by the exchange energy that is the most important term around the singularity.\cite{Doring-1968bf} Exchange interaction is responsible of ferromagnetic order and is the most divergent term around the center of the BP, giving the topological structure to BPs. In this paper we will concentrate in exchange energy that is a geometry-independent term , and so it can give us the universal results that can be considered as a first order contribution to spin wave dynamics. By performing a transformation of magnetization field into the complex plane we are able to calculate the spectrum of oscillations around the BP that turns to be the same as those of a quantum system of an electrical charge in a monopolar magnetic field. The mathematical analogy between Dirac monopole and spin waves dynamics around the BP allows us to calculate with ease the scattering of a SW in a BP field, opening a new possibility to BP detection and localization, by means of the observation of the interference pattern and the intensity profile of scattered spin waves. It is interesting to note that in some particular cases the same method give as result a Schr\"odinger-like equation for SWs; for example for single skyrmions,\cite{PhysRevB.75.132401} vortex domain walls \cite{Gonzalez2010530} and one-dimensional domain wall.\cite{PhysRevLett.107.177207} It is worth to note, however, that there are situations in which the equation for SWs are more complicated, giving sets of two coupled Schrödinger equation, as in the case of magnetic vortices. \cite{ PhysRevB.58.8464}

The paper is organized as follows: in the second section we present the physical system and the equations of motion, with the BP as a solution of them. In the third section we perform a change of variables into the complex plane and we perturb the equations of motion around the BP solution showing that the resulting linear equation is a Schr\"odinger-like equation for the interaction between an electron and a magnetic monopole. In the forth section we show the analytical solution for the oscillations and the functional form of the SWs. In the fifth section we study the scattering of a plane wave produced by the BP using the results of the previous sections and classical results on magnetic monopoles.

\begin{figure}
\centering
\hspace{0.3cm}$q=1,\gamma=0$\hspace{1.2cm}$q=1,\gamma=\pi/2$\hspace{1cm}$q=2,\gamma=0$\\
\includegraphics[width=0.15\textwidth]{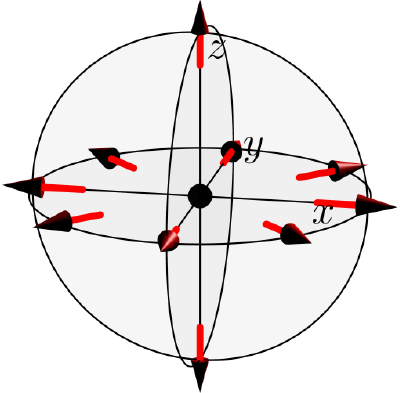}
\hspace{0.1cm}
\includegraphics[width=0.14\textwidth]{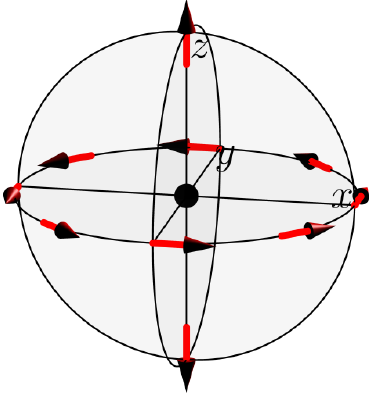}
\hspace{0.1cm}
\includegraphics[width=0.15\textwidth]{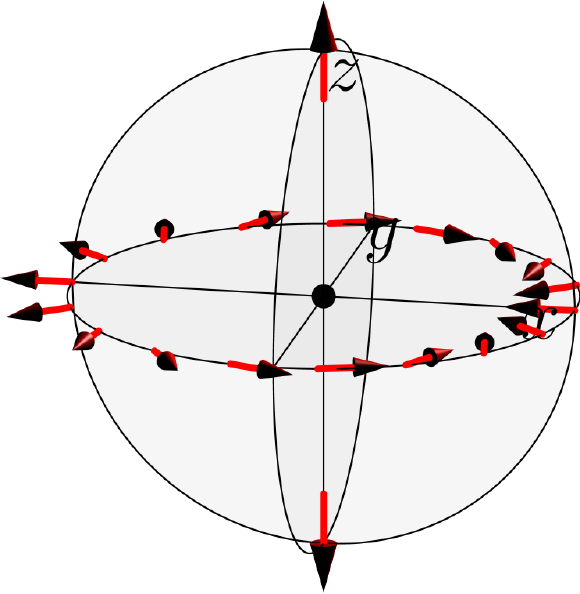}\\
\caption{Three kind of BPs depending on the vorticity $q$ and the phase in the azimuthal coordinate $\gamma$.}
\label{fig_1}
\end{figure}
 
\section{Model and Bloch point solution} 

We consider a system composed of dimensionless classical spins parameterized by field coordinates $(S,\Theta, \Phi)$ as $\bm S=S(\cos\Phi \sin\Theta, \sin \Phi \sin\Theta,\cos\Theta)$ where $\hbar S$ is the molecular spin proportional to the magnetization. The energy of the spin configuration is dominated, in the vicinity of the BP, by the exchange contribution ${\cal E}=\int({\rm d}V/2a) J(\nabla \bm S)^2$, with $J$ the exchange energy constant and $a$ the lattice parameter. In this frame, there is a natural cut-off for the wave-vectors norm corresponding to a minimum wavelength of the order of the exchange length $\ell_e$, defined by $\ell_e =\sqrt{J/{\cal M}_s^2 a}$, with ${\cal M}_s$ the saturation magnetization. This exchange length turn to be of around six nanometers in Permalloy. The kinetic term for such a system is the so called Berry term ${\cal S}_B=\int({\rm d}V/a^3) \hbar S\cos\Theta\dot\Phi$.
The magnetic texture predicted by this action acquires the form of a twisted BP and can be simply written as 
$
\Theta_{0}=p\theta$ and $
\Phi_0=q\phi+\gamma
$
where $\theta$ and $\phi$ are the usual spherical coordinates in space $\bm r=(r, \theta,\phi)$. Since the magnetization field is single valued we need $p$ and $q$ to be integers (see Fig. \ref{fig_1}). The free parameter $\gamma$ represents an azimuthal tilt with respect to the radial direction, making the BP twist around z. Its optimal value depends on additional terms in the energy, particularly the dipolar energy,\cite{Doring-1968bf} giving as result a twisted BP with a definite angle $\gamma$.  Works found in the literature\cite{Galkina-1993uf, Elias-2011fk, PhysRevB.86.094409} showed that if we allow the variation of the magnetization norm it is possible to estimate the size of the singular region for prototype magnet (for example permalloy); this singularity region is found to be of a few nanometers. The behavior of the magnetization in the vicinity of the BP is, nevertheless, fill with subtleties. Recent simulations reveal that to give a proper assessment of the the singular behavior it is necessary to resolve the magnetic degrees of freedom down to atomic resolution.\cite{PhysRevB.89.134403} This is consistent with some micromagnetic simulations that have shown that it is possible to stabilize a BP in a spherical domain of a few nanometers.\cite{Elias-2011fk,PhysRevB.85.224401} In this work we focus on the topological BP with topological charge  $Q=pq$ (also called the Pontryagin invariant that it is nothing but than the number of times that magnetization on a closed surface around the BP center covers the whole solid angle). In this article we will concentrate us in the case $p=1$ so the vorticity $q$ is the same as the topological charge $Q$. BPs (as vortices) cannot be considered localized solutions because the spin field is not homogeneous at infinite, and so, the BP energy $E_B$ calculated for a spherical domain is proportional to the radius of the sphere $R$ as $E_B=8JS^2Q(R/a)$.

\begin{figure}
\centering
\includegraphics[width=.2\textwidth]{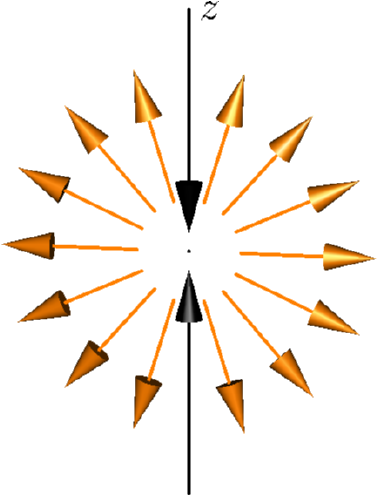}
\caption{The magnetic field $\bm B=q\frac{\hat{\bm r}}{r^2}-4\pi q\bm{\hat r}\delta(x)\delta(y)$ generated by the vector potential $\bm A(\bm r)=-\frac{q\cot\theta}{r}\hat \phi$. The black arrows are the singularity line where potential is infinity and field has opposite direction. In this case the singularity line is along the whole $z$ axis (this is called symmetric potential).}
\label{fig_2}
\end{figure}

\section{Spin waves excitations in the vicinity of a Bloch point} 

\begin{figure*}
\centering
${\cal Y}_{1-1}^{(1)}(\theta,\phi)$\hspace{1.3cm}${\cal Y}_{10}^{(1)}(\theta,\phi)$\hspace{1.3cm}${\cal Y}_{11}^{(1)}(\theta,\phi)$\\
\includegraphics[width=.15\textwidth]{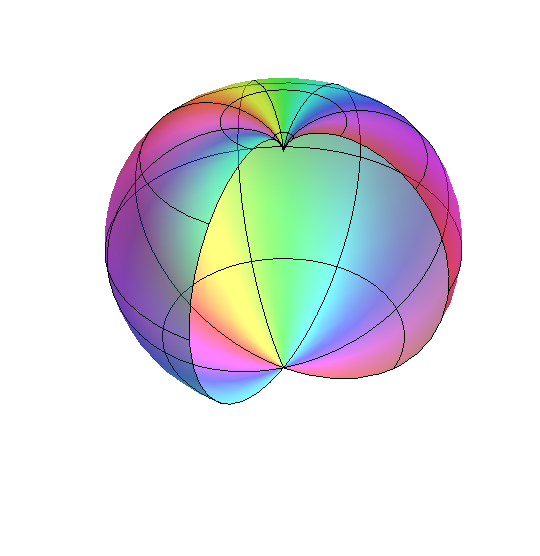}
\includegraphics[width=.15\textwidth]{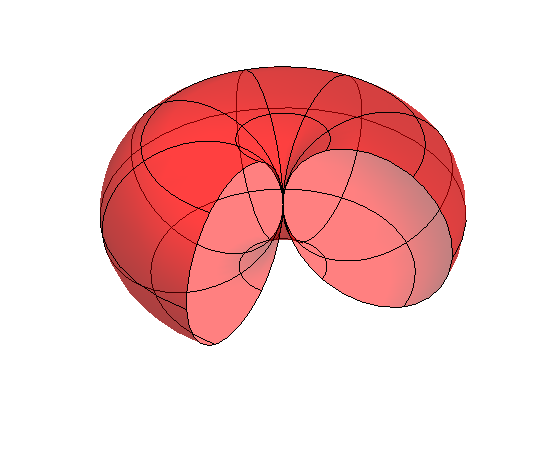}
\includegraphics[width=.15\textwidth]{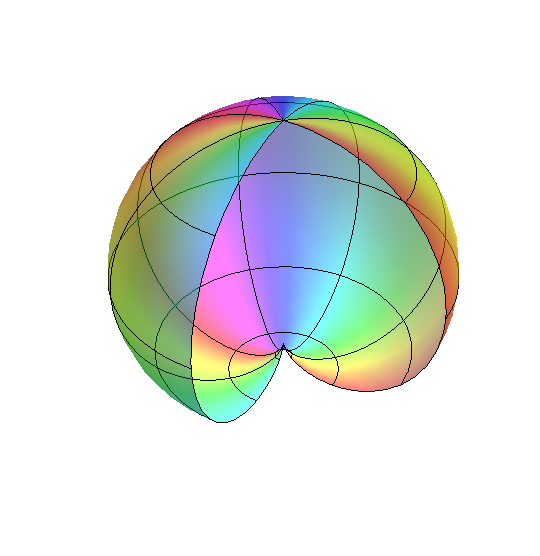}\\
\hspace{0.6cm}${\cal Y}_{2-2}^{(1)}(\theta,\phi)$\hspace{1.5cm}${\cal Y}_{2-1}^{(1)}(\theta,\phi)$\hspace{1.5cm}
${\cal Y}_{20}^{(1)}(\theta,\phi)$ \hspace{2.1cm}  ${\cal Y}_{21}^{(1)}(\theta,\phi)$\hspace{1.5cm} ${\cal Y}_{22}^{(1)}(\theta,\phi)$\\
\includegraphics[width=.157\textwidth]{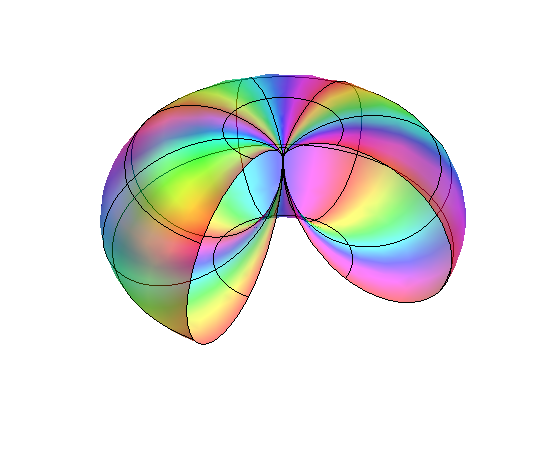}\includegraphics[width=.157\textwidth]{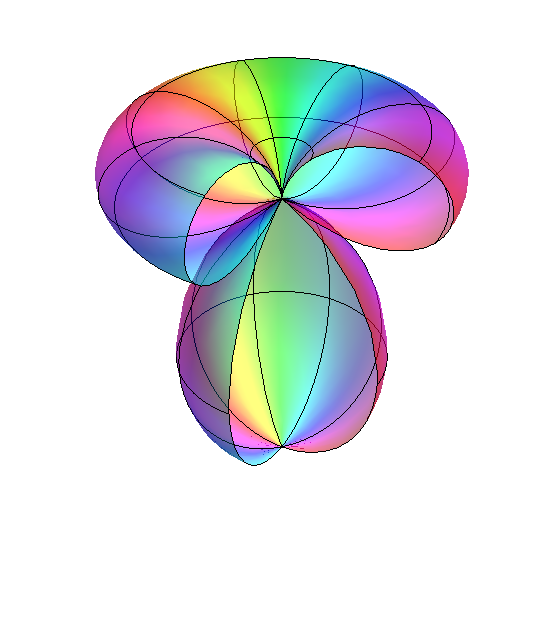}\includegraphics[width=.157\textwidth]{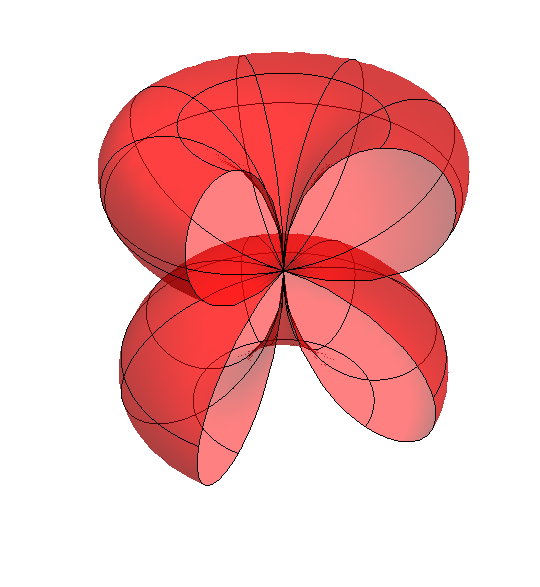}\includegraphics[width=.157\textwidth]{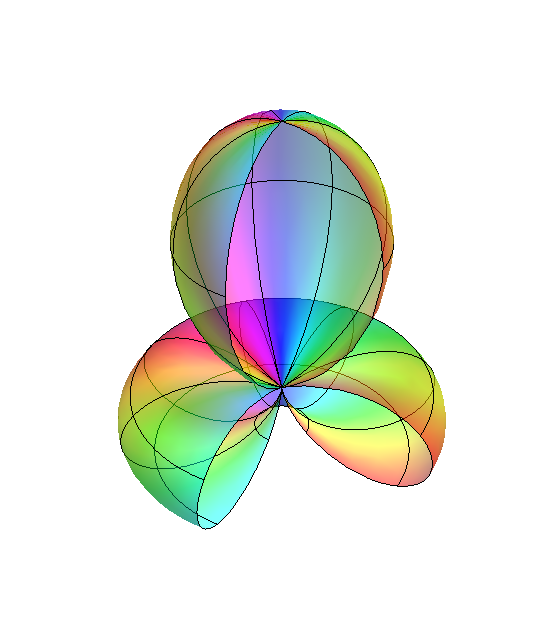}\includegraphics[width=.157\textwidth]{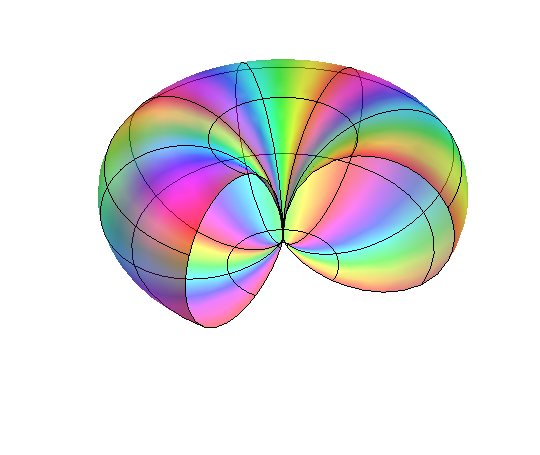}
\caption{Angular dependence of the eigenfunctions Eq. (\ref{solution_r}) for q=1, for different (l,m) modes; polar plot of the absolute value $|{\cal Y}_{lm}^{(q)}(\theta,\phi)|$ with, in color, their argument. }
\label{fig_3}
\end{figure*}

In order to calculate the SW excitations around the BP, we start from a given stationary solution parameterized by the spherical coordinates field $\Theta_0$ and $\Phi_0$ and consider a small distortion of the magnetization texture characterized by a local change $\delta\Theta$ and $\delta \Phi$. This distortion is readily associated with a change in the magnetization vector equal to $\delta \bm S=\delta\Theta\; \hat\Theta  + \sin\Theta_0  \delta\Phi\; \hat\Phi$. As expected the distortion lies in the tangent plane to the magnetization sphere, within this plane we follow  \cite{PhysRevB.75.132401} and use complex notation: 
$
\Psi=\delta\Theta+i\delta\Phi\sin\Theta_0.
$
The square of variations of the spin vector around a particular configuration are therefore related with the norm of  $\Psi$ by
$
\delta\bm S^2=S^2|\Psi|^2,
$
allowing us to interpret the "density of probability" $|\Psi|^2$ as the density of SWs. Expanding the magnetic action up until the second order in the perturbation $\Psi$ we obtain:
\begin{equation}
{\cal S}^{(2)}=\frac{1}{2}\int \overline\Psi\left(i\hbar \partial_t-\hat H_{B}\right)\Psi dtdV/a^3,
\end{equation}
with
\begin{equation}
\hat H_{B}=-\frac{\hbar^2}{2m^*}\left(\nabla^2+2iq\frac{\cot\theta}{r}\hat\phi\cdot\nabla-q^2\frac{\cos2\theta}{r^2\sin^2\theta}\right)
\end{equation}
where we have defined the equivalent mass $m^*=\hbar^2/4JSa^2$.

The equations of motion for the spin waves are obtained from the Euler-Lagrange equations of the linearized action,
$
\frac{\delta{\cal S}}{\delta \overline{\Psi}}=0
$,
giving
\begin{equation}
i\hbar\partial_t \Psi=\hat H_{B} \Psi.
\label{schro_gener}
\end{equation}
The effective Hamiltonian for the linear oscillations around the soliton can be written as $\hat H_{B}=\frac{\hbar^2}{2m^*}[-i\nabla +\bm A(\bm r)]^2+V(\bm r)$, 
$\bm A(\bm r)=\frac{q\cot\theta}{r}\hat \phi,
$ and $
V(\bm r)=-\frac{\hbar^2}{4m^*}\frac{ q^2}{r^2}.
$

The effective spin wave Hamiltonian is formally equivalent to that one describing a quantum mechanical charged particle moving under the influence of the magnetic field created by a magnetic monopole located at the singularity, and in a scalar attractive isotropic potential $V$. Away from the BP we have $\bm B=\nabla\times \bm A=-q\hat{\bm r}/r^2+4\pi q \hat{\bm r}\delta(x)\delta(y)$ (see Fig. \ref{fig_2}).  

The vector potential satisfies $\nabla\cdot \bm A=0$, the so called Coulomb gauge. The magnetic field corresponding to potential vector $\bm A$ is the famous Dirac monopole field \cite{Nakahara-2003fk,Shnir-2005uq} that is radial but having zero divergence (here and in what follows we adopt the formalism of Schwinger et al. \cite{Schwinger1976451}). The absence of a Coulomb term in the Hamiltonian means that the interaction is between two particles having one an electric charge and the other a magnetic charge, but not both kind of charges in the same particle (there are no ``dyons'' implied).

\section{Stationary solutions}
To solve this system we work in the standard way searching for the eigenvalues $E$ of the Hamiltonian $\hat H_B\Psi(\bm r)=E\Psi(\bm r)$. From the classical version of the magnetic  monopole problem it is known\cite{Schwinger1976451} that this Hamiltonian conserves a generalized version of the classical angular momentum $
\bm J=\bm r\times \bm p+q\hat{\bm r},
$ that in the quantum formalism give rise to the operators $\hat{\bm J^2}$ and $\hat{\bm J_z}$, which explicit form are
\begin{multline}
\hat{\bm J}^2=-\hbar^2\left[\frac{1}{\sin\theta}\frac{\partial}{\partial \theta}\left(\sin\theta\frac{\partial}{\partial \theta}\right)+\frac{1}{\sin^2\theta}\frac{\partial^2}{\partial\phi^2}\right]\\+2i\hbar^2q\frac{\cos\theta}{\sin^2\theta}\frac{\partial}{\partial\phi}+\frac{\hbar^2q^2}{\sin^2\theta}
\end{multline} 
\begin{equation}
\hat{\bm J}_3=-i\hbar\frac{\partial}{\partial\phi}.
\end{equation} 
Their eigenfunctions are given by the so called generalized spherical harmonics ${\cal Y}_{lm}^{(q)}(\theta,\phi)$ (see Appendix A). The eigenvalues problem is solved in a way similar to the standard angular momentum problem:
\begin{equation}
\hat{\bm J}^2{\cal Y}_{lm}^{(q)}(\theta,\phi)=l(l+1)\hbar^2{\cal Y}_{lm}^{(q)}(\theta,\phi),
\label{eq_eigen_1}
\end{equation} 
and
\begin{equation}
\hat{\bm J}_z{\cal Y}_{lm}^{(q)}(\theta,\phi)=m\hbar{\cal Y}_{lm}^{(q)}(\theta,\phi)
\label{eq_eigen_2}
\end{equation} 
where $l\geq |q|$, and $-l\geq m \geq l$ are integers. The generalized spherical harmonics share a series of properties with the usual spherical harmonics and have the important property of reducing to them when $q=0$, ${\cal Y}_{lm}^{(0)}(\theta,\phi)=Y_{lm}(\theta,\phi)$. 

Using the separation of variables $\Psi(\bm r)= {\cal R}(r){\cal Y}_{lm}^{(q)}(\theta,\phi)$ and defining 
$\ell(\ell+1)\equiv l(l+1)-2q^2
$, the radial differential equation becomes:
\begin{equation}
\left[\frac{1}{r^2}\frac{\partial}{\partial r}\left(r^2\frac{\partial}{\partial r}\right)+k^2-\frac{\ell(\ell+1)}{r^2}\right]{\cal R}(r)=0
\label{equation_r}
\end{equation}
where we have the dispersion relation $k^2=2m^*E/\hbar^2$ predicting a group velocity $\bm v=(2JSa^2/\hbar)\bm k$. This equation is solved by spherical Bessel function $j_\ell(kr)$ of order $\ell$.
With this we can write the general solution for the spin wave excitations in the form of an expansion:
\begin{equation}
\Psi(\bm r)=\sum_{l=q}^{\infty}\sum_{m=-l}^{l}A_{klm} j_\ell(k r){\cal Y}_{lm}^{(q)}(\theta,\phi).
\label{solution_r}
\end{equation}
The condition $l\geq q$ shows that there is only one mode of oscillation being zero at the origin: the one corresponding to $q=1$ and $l=1$ (see the relation previous to the Eq. (\ref{equation_r})), which gives a Bessel function of order $\ell=0$. In all the other cases the modes are different from zero at the origin. In this way, the singularity at $r=0$ is naturally avoid by most of the modes. We show in the Fig. (\ref{fig_3}) some representative modes of the angular part  of the solution. 

An important feature of these solutions is the absence of local modes. In fact, in our system there is no Coulomb-like potential, making that the radial part of the equation is equivalent to a free particle (aside the relation between the quantum numbers $\ell$ and $m$). This is closely related to the fact that, from the classical point of view, the angular momentum conservation in the interaction of the monopole and the electron gives an open trajectory over the surface of a cone in which the closest distance between the particles is the impact parameter. In general, solitons breaking the continuous translational symmetry should have special zero-modes that are usually quasi-local ones.\cite{Rajaraman-1987kx} These zero-modes do not appear in our treatment because we do not consider the rigid motion of the BP, fixing its position at the origin of coordinates.

\begin{figure}
\centering
\includegraphics[width=.4\textwidth]{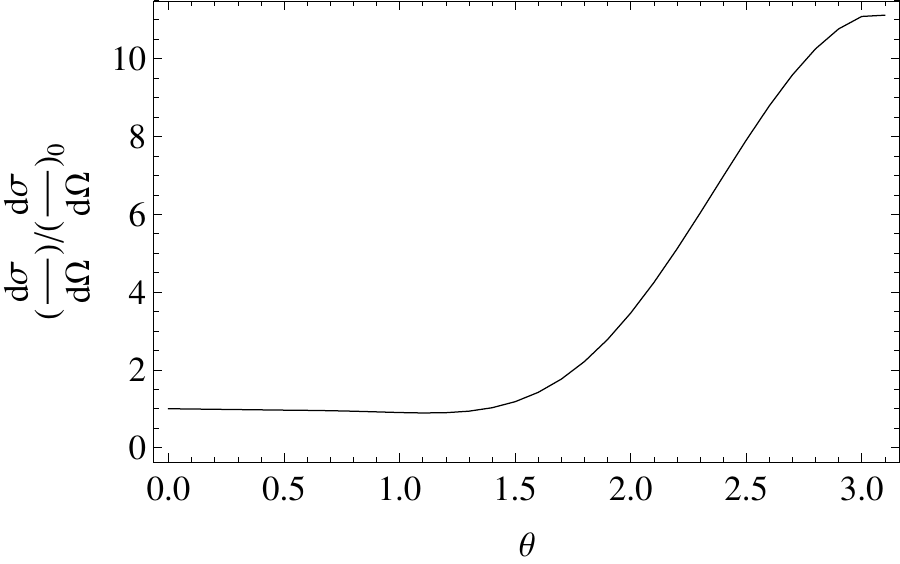}
\caption{Ratio of the differential cross section with respect to the small angles one, for q=1, as a function of the scattering angle.}
\label{fig_4}
\end{figure}

\section{Scattering of spin  waves by the Bloch point}

Let us now consider the problem of scattering of SWs by a BP. We have a straightforward relation between  the probability density $|\Psi|^2$ and the intensity of SWs.  Let us assume that we are far away enough from the BP in regions where magnetization is almost homogeneous.  There the Hamiltonian  reduces to a free particle whose simplest solutions are in the form of plane waves.  We can expand a planar wave into a series with the eigenfunctions of Eqs. (\ref{eq_eigen_1}) and (\ref{eq_eigen_2}).\cite{Schwinger1976451} Recognizing and setting apart the terms that represents an incident plane wave from those associated with an outgoing spherical wave we can find the amplitude of the spherical scattered wave $f(\theta)$, where $\theta$ is the angle between the incident wave and the point we are looking at (the scattering amplitude is independent of the choice of the singularity line). Explicitly  we have:\cite{Schwinger1976451}
\begin{multline}
2ikf(\theta)=\sum_{l=q}^\infty (2l+1)\mathcal{Y}^{(q)}_{lq}(\pi-\theta,\phi=0)e^{-i\pi\ell}.
\end{multline}
There is no explicit expression for this series and we need to evaluate it numerically. In the limit of small angles $\theta\ll 1$ the differential cross section becomes functionally equivalent to a Rutherford scattering:
\begin{equation}
\left(\frac{d\sigma}{d\Omega}\right)_0 = \left(\frac{q}{2k}\right)^2 \frac{1}{\sin^4(\theta/2)},
\end{equation}

We plot numerically the differential cross section divided by the small-angles limit, Fig. (\ref{fig_4}), for the topological charge $q=1$. 

The scattering of a spin wave moving across a BP singularity can be used to alter its location. Just like it has been proposed in the context of domain wall dynamics where  by means of the spin waves generated in the material\cite{Han-2009lo,PhysRevLett.107.177207} a force is found to over the domain wall. The result that we are presenting, concerning the behaviour of spin waves around BPs is of paramount importance to understand and to control BPs dynamics. 

\section{Discussion and conclusion}

In this paper we study a static and singular BP and the SWs in its presence. We use the simplest model that gives rises to a BP solution and its topological properties, that is, the exchange energy. By doing a very simple transformation that put the constant norm magnetization field into a complex variable, we are able to write a Schr\"odinger equation for the perturbations around the BP. This equation give us the dynamics of the SWs and it is found to be equivalent to the dynamics of the quantum interaction between an electron and a magnetic monopole, where the product of the electric charge and magnetic charge of the particles is the topological charge. We take advantage of the enormous understanding cumulated through the years on the subject\cite{Shnir-2005uq} to solve the Schr\"odinger equation (and found in this way the dynamics of the SWs) and to reinterpret the previous results on quantum scattering as the problem of free SW that find in their way a BP. The formula predict a Rutherford-like scattering for small angles, and a complex behavior for angles larger than $\pi/2$, especially when the topological charge of the BP is increased and we are close to the backscattering ($\theta\rightarrow \pi$). 

The properties of the scattering on a BP field open new possibilities for its detection by means of the study of its spectrum, and can be considered as the first order effects coming from the topology of the soliton. The difference in scattering of different charge BP can also be used to measure the topological charge. Analytical calculations on BP with other terms in the energy (as dipolar energy, anisotropies or external fields) are intrinsically complex and it is not yet clear what are the stability regions, even if it is commonly accepted that it is unstable in the presence of external field and anisotropies (considering form anisotropies and crystallographic ones). But even considering additional terms in energy, it is quite possible that the new situation follows the analogy between SWs in the BP field and the quantum interaction between an electron and a magnetic monopole because of topological reasons. The existence of a region with reduced magnetization rises a difficulty for the exact solution we are proposing. At a first glance the applicability of our solution is restricted to wavelengths larger than characteristic size of this region. New calculations must include modifications in the radial part that could support bound states, absent in our approach.  It could then be very interesting to test these kind of considerations in future works and the lost of stability of BPs by spin waves as well. 

\begin{acknowledgments}
R.G.E thanks Conicyt Pai/Concurso Nacional de Apoyo al Retorno de Investigadores/as desde el Extranjero Folio 821320024. V.L.C.S. thanks the Brazilian agency CNPq (Grant No. 229053/2013-0),
for financial support. The authors acknowledge funding from Proyecto Fondecyt numbers 11070008 and 1110271, Proyecto Basal FB0807-CEDENNA, Anillo de Ciencia y Tecnonolog\'ia ACT 1117, and by N\'ucleo Cient\'ifico Milenio P06022-F.
\end{acknowledgments}

\appendix
\label{appendix}
\section{Generalized spherical harmonics}
For completeness we show here the definition of the functions ${\cal Y}_{lm}^{(q)}(\theta,\phi)$. 

Giving the eigenfunctions of the operator $\hat{J}_z$ be $\Phi(\phi)=e^{im\phi}$ (with $m$ integer), the polar part of the operator $\hat{\bm J}^2$ gives the equation:
\begin{multline}
-\left[\frac{1}{\sin\theta}\frac{\partial}{\partial \theta}\left(\sin\theta\frac{\partial}{\partial \theta}\right)-\frac{m^2+2qm\cos\theta+q^2}{\sin^2\theta}\right]\bm \Theta\\=l(l+1)\bm \Theta.
\end{multline}
This equation is solved using the general rotation functions $U_{lm}^{(q)}(\theta)$. These functions are related to the Jacobi polynomials $P_{n}^{(b,c)}(x)$ (for $n$ integer) as:
\begin{multline}
U_{lm}^{(q)}(\theta)=\left[\frac{(l+q)!(l-q)!}{(l+m)!(l-m)!}\right]^{1/2}\times \\ \left(\frac{1-x}{2}\right)^{(q-m)/2}\left(\frac{1+x}{2}\right)^{(q+m)/2}P_{l-q}^{(q-m,q+m)}(x),
\label{def_Uqlm}
\end{multline}
where $x=\cos\theta$. The general rotations functions $U_{lm}^{(q)}(\theta)$ are used to define the generalized spherical harmonics
\begin{equation}
{\cal Y}_{lm}^{(q)}(\theta,\phi)\equiv \sqrt{2l+1}U^{(q)}_{lm}(\theta)e^{i m \phi},
\end{equation}
that are the eigenfunctions of the whole angular operator with the properties already mentioned in the text.

\bibliography{bp}

\end{document}